# Two-photon Induced Hot Electron Transfer to a Single Molecule in a Scanning Tunneling Microscope

S.W. Wu[*] and W. Ho[†]

*Department of Physics and Astronomy and Department of Chemistry,*

*University of California, Irvine, California 92697-4575, USA*

**Abstract:**

The junction of a scanning tunneling microscope (STM) operating in the tunneling regime was irradiated with femtosecond laser pulses. A photo-excited hot electron in the STM tip resonantly tunnels into an excited state of a single molecule on the surface, converting it from the neutral to the anion. The electron transfer rate depends quadratically on the incident laser power, suggesting a two-photon excitation process. This nonlinear optical process is further confirmed by the polarization measurement. Spatial dependence of the electron transfer rate exhibits atomic-scale variations. A two-pulse correlation experiment reveals the ultrafast dynamic nature of photo-induced charging process in the STM junction. Results from these experiments are important for understanding photo-induced interfacial charge transfer in many nanoscale inorganic-organic structures.

## I. INTRODUCTION

The understanding of photo-induced interfacial charge transfer in nanoscale inorganic-organic hybrid structures has recently been the focus of many research efforts, due to its fundamental importance and promising applications, such as efficient utilization of solar energy.[1-3] Of particular interest is the mechanism of such charge transfer processes at the interface. However, such information is often obscured in ensemble-averaged measurements because of the structural and electronic inhomogeneities at the interface.[1] To gain a better understanding of photo-induced charge transfer process, it is desirable to study single nanostructures in a well-characterized local environment.

The scanning tunneling microscope (STM) is a powerful tool that allows the study of single molecules on solid surfaces and the experimental determination of molecular configuration with atomic-scale resolution.[4] The ability to precisely position the STM tip over an individual molecule enables the investigation of electron transport through a single molecule in a metal-molecule-metal junction. This unique experimental configuration, in concert with theoretical efforts, has greatly improved our understanding of nanoscale electron transfer at inorganic-organic interface.[1-2, 5-6] In-situ laser irradiation of the metal-molecule-metal junction in a STM is expected to provide insight into photo-induced charge transfer in many nanoscale inorganic-organic hybrid structures.

In this paper, we report the study of photo-induced hot electron transfer by combining femtosecond optical excitation with a STM in the tunneling regime. The irradiation of femtosecond laser pulses to a STM junction leads to resonant tunneling of photo-induced hot electrons from the STM tip to molecular states of a single magnesium porphine (MgP) molecule adsorbed on a thin oxide film grown on NiAl(110) surface (Fig. 1). The photo-induced hot electron transfer, in this experiment, resulted in switching of the molecule from the neutral state to the anionic state, corresponding to a single molecule charging event. This approach of combining STM and laser is different from the irradiation-then-image method, where the STM tip is not in the tunneling regime during laser irradiation and the STM only serves as a surface characterization tool for imaging the products of photochemical reactions, such as dissociation, desorption and diffusion.[7-10] The in-situ laser irradiation of a STM junction enabled us to observe photo-induced processes in real time. Compared to CW lasers,[11] the irradiation of the STM junction in the tunneling regime with femtosecond laser pulses enables the study of nonlinear optical phenomena and potentially ultrafast dynamic processes in the STM junction.

## II. EXPERIMENTAL DETAILS

The experiments were conducted with a home-built low temperature STM housed in an ultrahigh-vacuum (UHV) chamber (Fig. 2),[12] providing a well-controlled experimental environment. The STM and sample were cooled by a continuous-flow cryostat using liquid helium and operated at a base temperature of 8.5 K in the absence of laser irradiation. The preparation and characterization of the sample followed the same procedures as described in our previous publications.[6, 11] Individual MgP molecules [Fig. 2(b)] were thermally sublimed onto a thin alumina film grown on NiAl(110) surface that has a thickness of ~0.5 nm. The STM tips were made of polycrystalline silver, due to its large electric field enhancement associated with surface plasmon resonance and lightning rod effect.

The STM junction was irradiated by a femtosecond Ti:Sapphire laser, which was operated at a repetition rate of 90 MHz with the central wavelength of about 807 nm. The pulse duration was ~70 fs as measured with a second-order autocorrelator consisting of a beta barium borate crystal before the laser entered the chamber. To align the laser to the junction, a pair of spherical lenses was mounted on two separate translation-and-tilt stages inside the UHV chamber. The laser incident and exit angles are 45° from the surface normal. One of the lenses was used to focus the laser into the junction; the other was used to image the STM junction onto a charge-coupled device (CCD) camera. When the laser was aligned to the junction, the STM tip was shown on the bottom in Figure 2(c) and its mirror image appeared on the top, due to the high reflectivity of the sample. The focal spot in the junction was estimated to be 20~50 μm in diameter. The laser electric field was linearly polarized in the plane of incidence (p-polarized), except for the polarization-dependent measurements. For comparison, a continuous wave (CW) HeNe laser (633 nm) was also used to irradiate the STM junction in the control experiments.[11]

Laser irradiation of the junction caused the temperature in the STM to rise linearly with the incident power at a rate that varied from tip to tip (0.2~0.7 K/mW), due to the different tip shape and shadow effect. The experimental data were acquired after the STM reached a new thermal equilibrium under laser irradiation. However, the thermal fluctuation associated with irradiation could induce oscillation in the tip-sample distance. Normally, the power of a laser fluctuates on the order of 1%. To reduce the power fluctuation, a laser power controller (LPC, Model BEOC-LPC, Brockton Electro-Optics

Corp., Brockton, MA) was used. The LPC effectively reduces the power instability by more than one order of magnitude [Fig. 2(d)]. The thermal fluctuation was further minimized by limiting the incident laser power to below ~2 mW. Because of the high repetition rate (90 MHz) of the femtosecond laser, energy of each pulse is less than ~22 pJ. By taking these measures, the power fluctuation of the incident laser to the junction is less than 1 µW. The STM junction under laser irradiation was stable enough, even allowing us to conduct tunneling spectroscopy with the STM feedback turned off.

The combination of a STM with femtosecond optical excitation, in principle, would enable the study of dynamic processes in the tunnel junction. A two-pulse correlation experiment was conducted to study photo-induced charging as a function of pulse delay. In the two-pulse correlation experiment [Fig. 2(a)], a collinear configuration was used, which allowed easier determination of the time zero between the two pulse trains overlapping in the STM junction as measured outside the UHV chamber with the autocorrelator. The two trains of laser pulses were set to equal power by using a 50/50 beamsplitter and neutral density filters. The time delay between the two trains was controlled by a stepping motor in one arm of the Michelson interferometer. It should be noted that this two-pulse correlation setup could be easily converted into one femtosecond pulse train by blocking one arm of the Michelson interferometer. In the polarization-dependent measurement with one pulse train, the rotation of femtosecond laser polarization was achieved by rotating a quarter-wave plate inserted in one arm of the Michelson interferometer. Because of the round trip of laser pulses in the arm, the quarter-wave plate effectively acts as one half-wave plate for rotating the linear polarization of a laser.

The collinear configuration inevitably induced interference, when the two pulse trains were temporally overlapped within the pulse duration. To average the measured quantity such as the charging probability, a small sine-wave modulation (~130 Hz) was applied to the time delay by shaking a piezo actuator in the other arm. The modulation amplitude was adjusted to $\lambda/4$ (~ 0.2 μm) so that the time delay (peak to peak) was modulated at $\lambda/c$, which corresponds to one optical cycle. When the two pulses were temporally overlapped within the pulse duration, they interfered constructively and destructively within about one modulation period. However, the thermal effect in the STM junction, in principle, was able to follow the constructive and destructive interferences at the modulation frequency (~130 Hz). Therefore, a small laser power was used so that the temperature fluctuation at the STM junction was only several tens of mK, and the STM is still stable enough to conduct the experiments described in this paper. When the delay was beyond the temporal overlap, the temperature of the STM remained steady, because laser irradiation with its power stabilized by the LPC was acting as a steady thermal load on the STM junction.

### III. RESULTS AND DISCUSSION

#### A. Photo-induced charging of a single molecule

In the absence of light, the charging of the molecule from the neutral to the anion occurs at positive sample bias by filling the lowest unoccupied molecular orbital, LUMO-$\alpha$ with one extra electron. The threshold of charging corresponds to the onset of LUMO-$\alpha$ state (~0.55 V, the difference from molecule to molecule is less than 0.1 V.). Laser irradiation to the single-molecule junction opens additional electron transfer channels, tunneling of photo-induced hot electrons into the LUMO+1 and higher orbitals, even when the bias voltage is set below the onset of LUMO-α [Fig. 1(b)]. The effective barrier height is lowered for tunneling into higher lying molecular states, which dramatically increases the tunneling probability. For example, with HeNe CW laser irradiation (λ = 633 nm, photon energy of 1.96 eV), the voltage threshold of charging is ~0.15 V,[11] which corresponds to one-photon energy below the onset of LUMO+1 at ~2.1 eV (The value was determined from tunneling electron induced fluorescence spectra[13] and density functional theory (DFT) calculation[14].).

The irradiation of femtosecond laser to the STM junction enables the charging of the molecule below the voltage threshold based on one-photon excitation with CW lasers. To quantify this photo-induced charging process, we used the experimental scheme as described in Ref. 11. This scheme is implemented by repeating the charging-probing cycles [Fig. 3(a)]. Every cycle starts with the charging period, in which the STM tip is positioned over one location of a neutral molecule at a chosen sample bias voltage $V_c$ and tunneling current $I_c$ for a fixed holding time $T_{hold}$. During this period, the molecule has the possibility of being charged with one extra electron by the tunneling of photo-excited hot electrons. The charge state of the molecule is probed by ramping the sample bias voltage negative. If the molecule is charged, the tip will retract farther away (larger $\Delta Z$) from the sample due to the emergence of the singly occupied molecular state. The molecule is discharged at certain negative bias voltage, marked by arrows in Fig. 3(a), and reset to the initial neutral state.

A real-time trace of several successive charging-probing cycles is shown in Fig. 3(a). The result of every cycle could be simplified as either charged or uncharged. The bin time per cycle is determined by the holding time $T_{hold}$ for the charging period. With this method, the time trajectory of photo-induced charging events was recorded over a long period of time at one specific condition including fixed laser power and charging bias voltage $V_c$. One example is shown in Fig. 3(b). The individual black bar represents a single charging event. The time trajectory typically took 1-2 days to accumulate more than 800 charging events by maintaining the STM tip over the same position of the molecule. During the entire acquisition process, the molecule remained the same as checked by dI/dV spectra and STM topographic images.

The long trajectories permit detailed statistical analysis. The most straightforward analysis is to plot the distribution of the time interval between two consecutive charging events, which corresponds to the "waiting time" for the molecule to be charged. This approach is similar to the analysis of on- and off-times for the intermittent fluorescence intensity trajectories, which reveals single molecule dynamics such as intersystem crossing and chemical reactions.[15, 16] However, in our case, the discharging of the molecule is not photo-induced and is achieved by regulating the bias voltage. In Fig. 3(c), the histogram of time interval derived from the time trajectory shown in Fig. 3(b) is plotted, revealing the dominance by a single exponential decay. The exponential distribution in the "waiting time" analysis suggests that the photo-induced charging of the molecule is a Poisson process and the charging rate is characterized by $1/\tau$, where $\tau$ is the decay constant obtained from the exponential distribution. The confirmation of Poisson process justifies that the photo-induced charging could also be characterized by the charging probability p, a statistical mean value given by the number of charging events over the total number of cycles, n. The error bar of charging probability is given by $\sqrt{p(1-p)/n}$. In our experiment, the use of charging probability is more efficient due to reduced data acquisition time (we used 150 cycles to obtain one statistical mean value p.).

### B. Two-photon excitation by femtosecond laser pulses

The charging probability as a function of sample bias voltage $V_c$ under femtosecond laser irradiation shows different behavior to that with CW lasers [Fig. 4(a)]. Although the one-photon threshold of $V_c$ (~0.5 V) for CW, 800 nm irradiation is still evident, the charging probability is non-zero below that threshold, even at negative sample bias. Furthermore, the charging probability is higher at positive bias than at negative bias. Because of the short time duration of femtosecond laser pulses, the peak power is much higher (a factor of ~$1.6\times10^5$) than a CW laser with the same average power. The high peak power momentarily induces a large number of hot electrons, possibly resulting in resonant tunneling of hot electrons into the LUMO state instead of LUMO+1. The lower threshold of $V_c$ may also result from resonant tunneling of higher energy hot electrons via multiphoton excitation to higher lying molecular states.

To determine which mechanism is responsible for charging below the one-photon threshold, the charging probability p and charging rate ($1/\tau$) as a function of the incident power P are obtained and shown in Fig. 4(b) and Fig. 4(c). The bias voltage $V_c$ was set below the one-photon threshold when the femtosecond laser pulses illuminated the junction. The data were fitted to power law, $p \propto P^\gamma$ or $1/\tau \propto P^\gamma$, yielding $\gamma = 2.14 \pm 0.28$ from Fig. 4(b) and $\gamma = 2.01 \pm 0.05$ from Fig. 4(c). In comparison, illumination by CW laser yields the linear dependence [$\gamma = 0.96 \pm 0.13$ from Fig. 4(b) and $\gamma = 1.04 \pm 0.07$ from Fig. 4(c)] on the incident power. The quadratic dependences of the charging probability and charging rate on the incident power suggest two-photon excitation by femtosecond laser pulses. The exponent of γ ~ 2 with femtosecond optical excitation rules out mechanisms based on one-photon excitation, such as resonant tunneling to the LUMO state instead of LUMO+1 due to momentary increased number of hot electrons in the tip.

The two-photon excitation mechanism is also substantiated by the polarization-dependent measurement of the charging probability (Fig. 5). It is found that the photo-induced charging sensitively depends on how the incident laser electric field is oriented with respect to the STM junction. The excitation in the tip is most efficient when the incident laser is p-polarized, where half of the power could be projected into the direction along tip axis. The ratio of normalized charging probability between p-polarized and s-polarized light is very large, ~$10^3$ in the two-photon absorption regime with femtosecond laser; ~30 for CW laser in the case of one-photon excitation. Because of this large ratio, the data could be simply fitted by $\cos^2(\theta)$ for one-photon excitation or $\cos^4(\theta)$ for two-photon excitation, where θ = 0° or 180° for p-polarized light and θ = 90° or 270° for s-polarized light. The function of $\cos^2(\theta)$ represents the projection of laser power into the plane containing the incident laser direction and tip axis; while $\cos^4(\theta)$ reflects the square of the laser power projected onto that plane for the two-photon excitation. The data were also fitted by considering the incident angle of 45° from the surface normal, but the difference between the two fitting approaches is negligible. The excitation preference for p-polarized light is due to the larger electric field enhancement along the tip axis, originating from the dipole-like antenna formed by the tip-sample junction.[17]

Although the power and polarization dependences favor the two-photon process, the hot electrons photo-excited by one-photon absorption are still able to reach the LUMO state, and the number of these hot electrons is larger than that excited by two-photon absorption. The two-photon dominance can't be simply explained by the high peak power of femtosecond laser pulse. Rather, the effective tunneling barrier height plays an important role. The effective tunneling barrier height is one of the key parameters in determining the tunneling probability of electrons from the STM tip to the molecular states. In the simplest form, $T \propto e^{-A\sqrt{\Phi}\Delta z}$, where T is the tunneling probability, $A = 2\sqrt{2m}/\hbar$, $\Phi$ is the effective barrier height, and $\Delta z$ is the barrier width (distance between the tip and molecule). For an unbiased tunneling junction, the effective barrier height normally depends on the work functions of the tip and sample. Experimentally, we have measured the effective barrier height by recording an I-Z curve over a molecule.[6] The effective height for the molecule in neutral state is about 3.54 eV, smaller than the work function of silver tip (4.6 eV) due to the image potential effect and other factors.[18] For photo-excited hot electrons in the tip, the effective barrier height is much lower than that for normal electrons at the Fermi level in the tip, depending on the photon energy. When the STM junction is irradiated by femtosecond laser pulses (807 nm), the tunneling probability for two-photon excited hot electrons in the tip is about three orders higher than that for normal electrons at the Fermi level, and about two orders higher than that for one-photon excited hot electrons. Therefore, the two-photon excited hot electrons have the highest probability to resonantly tunnel into the molecular states and result in molecular charging, despite the smaller number than that of one-photon excited hot electrons. Furthermore, the two-photon excited hot electrons are more likely to find resonant molecular states to tunnel, because the electronic states are much denser as the energy increases. The excess energy associated with the decay from the LUMO+1 or higher lying molecular states to the LUMO could be dissipated in the radiative and non-radiative transitions and in the reorganization energy of the oxide film in the vicinity of the negatively charged molecule.

The effective tunneling barrier height not only determines the tunneling probability of hot electrons but also explains the disparity of charging probability at positive and negative bias voltages. At positive bias, the photo-induced electrons tunnel in the same direction as the continuously flowing tunneling electrons [Fig. 6(a)]. Assuming a rectangular barrier shape when unbiased, the barrier height $\Phi$ decreases on the average by $|V_b|/2$ and becomes $\Phi = W - 2h\nu - |V_b|/2$, where W is the barrier height for normal electrons at the Fermi level.[19] On the contrary, the photo-excited hot electrons in the tip flow against the DC electric field in the junction at negative bias [Fig. 6(b)]. Instead, the barrier height increases by $|V_b|/2$, rendering $\Phi = W - 2h\nu + |V_b|/2$. Because the tunneling probability exponentially depends on the square root of the effective barrier height, the charging probability at negative $|V_b|$ is always lower than at positive $|V_b|$ below the one-photon threshold, as shown in Fig. 4(a).

The femtosecond laser irradiation leads to the possibility that the molecule is charged by resonant tunneling of hot electrons from the substrate, instead of the STM tip. However, that contribution is very small, as indicated by the spatial dependent measurement of charging probability (Fig. 7). The photo-induced charging probability by the in-situ irradiation of femtosecond laser was measured along a line through the two-lobe structure of a magnesium porphine (MgP) molecule, shown as red line-connected filled squares. The junction was illuminated by a femtosecond laser at P = 84 µW, and the sample bias voltage $V_c$ was set at 0.3 V. For comparison, the apparent heights cut along the same line from constant current STM images Fig. 7(b) and (c) are also shown in black dashed curve for $V_b$ = 2.0 V and in black dot-dashed curve for $V_b$ = -0.3 V. When the bias voltage $V_b$ is set above the unoccupied states for molecules adsorbed on a thin insulating film, the apparent height in STM images is dominated by the electronic structure of the molecule.[6, 20] If the bias voltage $V_b$ is set below the onset of molecular states, the apparent height reflects its atomic structure. From the data, the photo-induced charging probability spatially varies from one location to the other within the molecular electronic contour. Outside the electronic contour, the charging probability is close to zero. The non-zero charging probability outside the molecule is possibly due to the tunneling of photo-induced hot electrons from the substrate. In addition, it can also arise from charging when the STM tip is positioned on the molecule during tracking, as one of steps in the experimental scheme described in Ref 11. Therefore, the photo-induced charging probability by femtosecond laser pulses is dominated by the tunneling of hot electrons from the STM tip. This reflects that the excitation of hot electrons in the silver tip is more efficient than the NiAl substrate because of the enhanced laser electric field near the tip apex.[17]

Because photons are coupled to the tunneling process, the spatial resolution is limited by the spatial confinement of tunneling electrons and the electronic contour of the molecular states on the surface. The electronic states in the molecule are more delocalized than the tunneling electrons. The spatial variation in charging probability by femtosecond laser does not change compared with that obtained from one-photon excitation by CW lasers.[11] This result is distinguished from the spatial resolution obtained only with local electric field enhancement, where two-photon excitation leads to a significantly improved spatial contrast over the one-photon excitation.[21]

### C. A route to probe ultrafast dynamics with the STM

The combination of a STM with femtosecond optical excitation, in principle, would enable the study of ultrafast dynamic processes in the tunnel junction.[22] A few efforts have been devoted towards this goal, including the detection of surface photovoltage,[23-28] photoemitted electrons,[29] surface plasmons,[30] and adsorbate desorption.[31] However, some of the studies were not actually conducted in the tunneling regime when the STM junction was irradiated with femtosecond laser pulses; some were severely linked with thermal effects due to irradiation of laser pulses. Our experimental results clearly identified the photo-induced hot electron tunneling process in a STM combined with femtosecond optical excitation. To further elucidate the dynamic aspect of the photo-induced charging process, a two-pulse correlation experiment was carried out.

In the two-pulse correlation experiment, the charging probability was measured as a function of the time delay between the two pulse trains. Figure 8 shows the data of the two-pulse correlation experiment, when the STM tip was positioned on one location of a molecule. The charging probability reaches the highest when the two pulse trains are overlapped. As the time delay varies away from the pulse overlap, the charging probability decreases and its envelop decays within a few hundred fs, which is noticeably longer than the pulse duration. Furthermore, the data barely, but reproducibly show a few equally-spaced oscillations around 125 fs, 250 fs and 375 fs at both positive and negative delays (guided by the grid lines in Fig. 8), despite the relatively large error bars in the data. When the two trains are separated with time delay greater than 600 fs, the charging probability stays constant within the error bars (not shown). The value of this charging probability is twice as large as that of charging probability when only one of the two equally powered pulse trains illuminates the junction. The two-pulse correlation measurements were also conducted on a different part of the molecule and another molecule with a different STM tip, and the dynamic behavior remains the same.

In the two-pulse correlation experiments, the charging probability is a measure of the rate of photo-induced hot electron tunneling. The first femtosecond laser pulse in each pulse pair impulsively excites the STM junction, and creates a burst of hot electrons in the STM tip via either one-photon or two-photon absorption. These initially created hot electrons follow a highly nonthermal distribution and have energies well above the Fermi level (up to 3.1 eV). Note that the ballistic transport time is 1-10 fs for hot electrons within the mean free path to the tip apex and the traveling time is on the sub-fs scale for electron tunneling.[18] One of these two-photon excited hot electrons could possibly resonantly tunnel into the molecular states on the sample, thus contributing to the rate of photo-induced hot electron tunneling. Otherwise, all the photo-excited hot electrons decay via various interactions such as electron-electron and electron-phonon scatterings. Those interactions affect the population and lifetime of hot electrons when the second pulse arrives after a tunable time delay. Therefore, the charging probability as a function of pulse delay time can reflect such interactions involving the hot electrons in the tip. The decay time of a few hundred fs observed in the two-pulse correlation experiment is close to the time scale of electron-electron interactions in silver (~350 fs).[32] This agreement is understandable because the initially created hot electrons have lost most of their excessive energy in the first relaxation step due to electron-electron scatterings.

The oscillations at 125, 250, and 375 fs, discernible in both positive and negative delays, suggest an additional coherent dynamics in the photo-induced charging process. Such coherent dynamics could possibly arise from coherently excited phonon modes in the STM tip by the first laser pulse that affect hot electron population via electron-phonon coupling upon the second laser pulse illumination.[33] The coherence time of ~125 fs corresponds to a mode of ~8 THz or ~33 meV. However, the energy of phonons in silver does not exceed to 5 THz.[34] Therefore, the observed oscillation could not be simply explained by the electron-phonon interaction in the STM tip. We speculate that the observed coherent oscillation possibly arises from coherently excited molecular vibrational mode of MgP[35] or phonons in the NiAl oxide substrate[36, 37], in which the mode of ~8 THz or ~33 meV could be accounted. In the two-pulse correlation experiments, the charging probability depends not only on the photo-induced hot electron population, but also on the instantaneous physical properties of the molecule on the sample, which affects stabilization of the anionic molecule after gaining one extra tunneling electron. The first femtosecond laser pulse impulsively excites electrons in the tip and coherently excites vibration in MgP or phonons in NiAl oxide substrate. The photo-induced charging by the delayed second laser pulse could be affected by this coherently excited vibrational mode in the molecule or phonons in NiAl oxide substrate. And this special mode of ~33 meV may be exactly responsible for the stabilization of anionic molecule on the oxide surface. We emphasize here that the above proposed mechanism is just a possible scenario and confirmation of such mechanism needs further theoretical and experimental investigations, which is beyond the scope of this paper.

## IV. CONCLUSION

In summary, we presented a novel approach to the study photo-induced hot electron transfer to a single molecule by in-situ irradiation of femtosecond laser pulses to a STM junction. The tunneling of two-photon excited hot electrons from the STM tip to the molecule was found responsible for the single-molecule charging, in which the much reduced effective tunneling barrier for photo-excited electrons plays a significant role. The approach and the results are important for understanding photo-induced charge transfer in inorganic-organic interfacial nanostructures such as passivated semiconductor nanocrystals on $TiO_2$ surface,[38] which is pivotal to the development of solar energy conversion and molecular electronics.[1-3] This novel approach could also lead to the study of dynamic processes with atomic scale resolution in real space by exploiting the temporal aspect of femtosecond lasers.[22]

**ACKNOWLEDGEMENTS**

This material is based on work supported by the Chemical Science, Geo- and Bioscience Division, Office of Science, U.S. Department of Energy, under grant DE-FG02-04ER1595.

# References

*Present address: Molecular Foundry, Lawrence Berkeley National Laboratory, Berkeley, CA 94720, USA.

†wilsonho@uci.edu

# Figure Captions

**Fig. 1.** (color online) Photo-induced charging on a single molecule. (a) A single MgP molecule in the STM junction under femtosecond laser irradiation. The photo-excited electrons (red dots) in the STM tip tunnel into the molecule, converting it from the neutral to the anion - charging. (b) Energy diagram showing the mechanism of photo-induced charging via one-photon and two-photon excitation.

**Fig. 2.** (color online) Experimental setup. (a) Schematic of optical path. LPC: laser power controller, BS: beamsplitter, ND: neutral density filter, HWP: half-wave plate, QWP: quarter-wave plate. (b) Molecular structure of magnesium porphine (MgP). (c) A CCD image showing the STM junction when the laser was aligned. The bottom one is the STM tip and the top one is its mirror image, due to high reflectivity of the sample. (d) Power of the femtosecond laser recorded over several hours with (red curve) and without the LPC (black curve). With the LPC, the power fluctuation is reduced by 23 times.

**Fig. 3.** (color online) Time trajectory and histogram of photo-induced charging events. (a) Real-time trace of several successive charging-probing cycles under illumination of a femtosecond laser ($\lambda$ = 807 nm, $V_c$ = 0.3 V). The time periods (5s) to attempt charging are shaded with gray bar. Detailed description of the measurements is discussed in the text. (b) A time trajectory of photo-induced charging events was recorded at the power level of 69 µW. For simplicity, those charging-probing cycles resulting in photo-induced charging events are represented with individual black bars. The bin time per cycle is 5 seconds, the length of time during which charging can occur. (c) Histogram of the time

interval between neighboring charging events in a semi-log plot. The data were fitted with a function of Aexp(-t/τ), yielding the charging rate (1/τ) of $0.062 \pm 0.002$ s$^{-1}$.

**Fig. 4.** (color online) Two-photon excitation by femtosecond laser pulses. (a) Charging probability as a function of sample bias voltage $V_c$: under the illumination of a femtosecond laser (λ = 807 nm, 34 μW, red open diamonds), CW lasers (λ = 800 nm, 146 μW, magenta filled triangles and dot-dashed line; and λ = 633 nm, 7.4 μW, green filled circles and dashed line), and without laser illumination (black crosses). The data with CW laser illumination were taken in different experimental runs. The dashed and dot-dashed lines yield the threshold of one-photon excitation. The inset shows the tip height displacement ΔZ while ramping the sample bias $V_b$ with the feedback on (current setpoint = 50 pA). The arrow marks the onset of LUMO-α of the neutral molecule. The decrease in the tunneling gap (corresponding to a decrease in ΔZ) can account for the increase in the charging probability from 0.45 V to 0.2 V. Charging probability (b) and charging rate (c) as a function of incident power under illumination of a femtosecond laser (λ = 807 nm, $V_c$ = 0.3 V, red open diamonds) and a CW laser (λ = 633 nm, $V_c$ = 0.2 V, green filled circles). The sample bias voltage $V_c$ were set below one-photon threshold for the femtosecond laser and above one-photon threshold for the CW laser. The dependences on the incident laser power were fitted to power law, $p \propto P^\gamma$ or $1/\tau \propto P^\gamma$. For the femtosecond laser, γ = 2.14 ± 0.28 (b) and γ = 2.01 ± 0.05 (c); for CW laser, γ = 0.96 ± 0.13 (b) and γ = 1.04 ± 0.07 (c).

**Fig. 5.** (color online) Laser polarization angle dependence of charging probability. The charging probabilities were measured as the linear polarization of the laser field was rotated. The angle of incidence is 45° from the surface normal. The p-polarized light is defined as 0° or 180°; the s-polarized light is defined as 90° or 270° (A difference in angle of up to 20° between the laser incident plane and the tip axis, which originates from the imperfectness of tip axis alignment and nonsymmetrical shape of the tip near its apex,[39] has been observed and corrected in our experiment.) The radial axis is shown in a log scale in order to enlarge the data contrast. The data are normalized by dividing the charging probability by the laser power for CW laser illumination ($\lambda$ = 633 nm, $V_c$ = 0.3 V, green filled circles) or square of the laser power for femtosecond laser illumination ($\lambda$ = 807 nm, $V_c$ = 0.3 V, red open diamonds). The normalized charging probabilities were fitted by $\cos^2(\theta)$ (dash line) or $\cos^4(\theta)$ (solid line), depending on one-photon or two-photon excitation.

**Fig. 6.** (color online) The effective barrier height for photo-induced hot electrons to tunnel from the STM tip to the molecule under positive (a) and negative (b) sample bias, respectively.

**Fig. 7.** (color online) Atomic-scale spatial variation of photo-induced charging probability. (a) The charging probabilities were measured along a line through the two-lobe structure of the molecule, shown as red line-connected filled squares. The junction was illuminated by a femtosecond laser at P = 84 µW, and the sample bias voltage $V_c$ was set at 0.3 V. The apparent heights cut along the same line from STM images (b) and

(c) are also shown in black dashed curve for $V_b = 2.0$ V and in black dot-dashed curve for $V_b = -0.3$ V.

**Fig. 8.** (color online) Charging probability as a function of time delay in two-pulse correlation measurement. Three sets of data measured with tip positioned on the bigger lobe of the molecule are shown here in red squares (1.8 pJ/pulse), green circles (1.2 pJ/pulse) and blue triangles (1.2 pJ/pulse). Although all three sets of data were measured on the same molecule with the same tip, laser alignment to the STM junction varied in each run. The sample bias $V_c$ was set at 0.3 V. For comparison, the second-order autocorrelation data in a collinear setup are also shown in black dots. The femtosecond laser pulses were slightly chirped because of the thick glasses used in the optical path such as the LPC laser power stabilizer and UHV view ports.

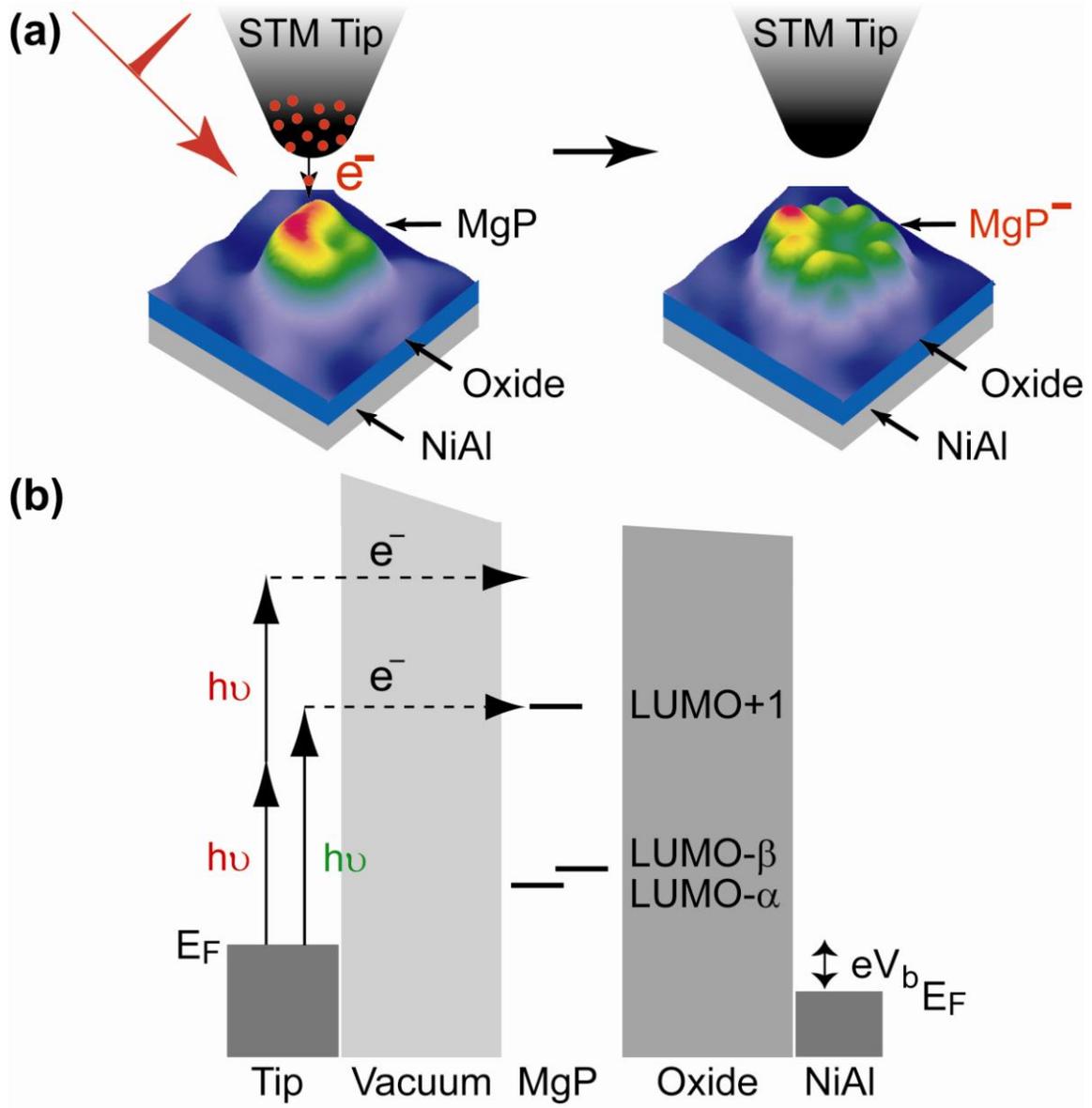

Figure 1

Figure 2

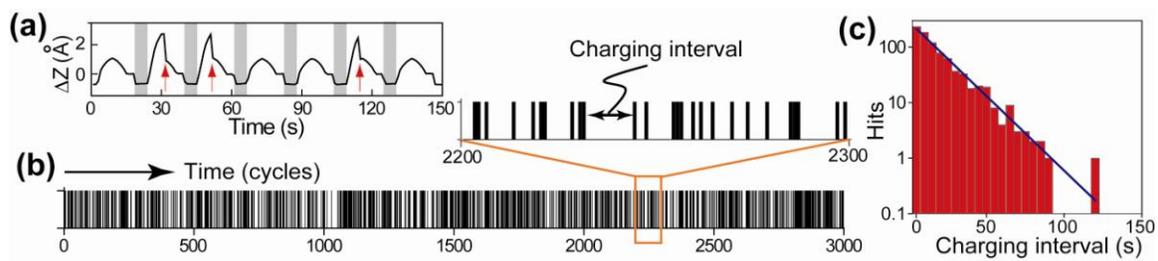

Figure 3

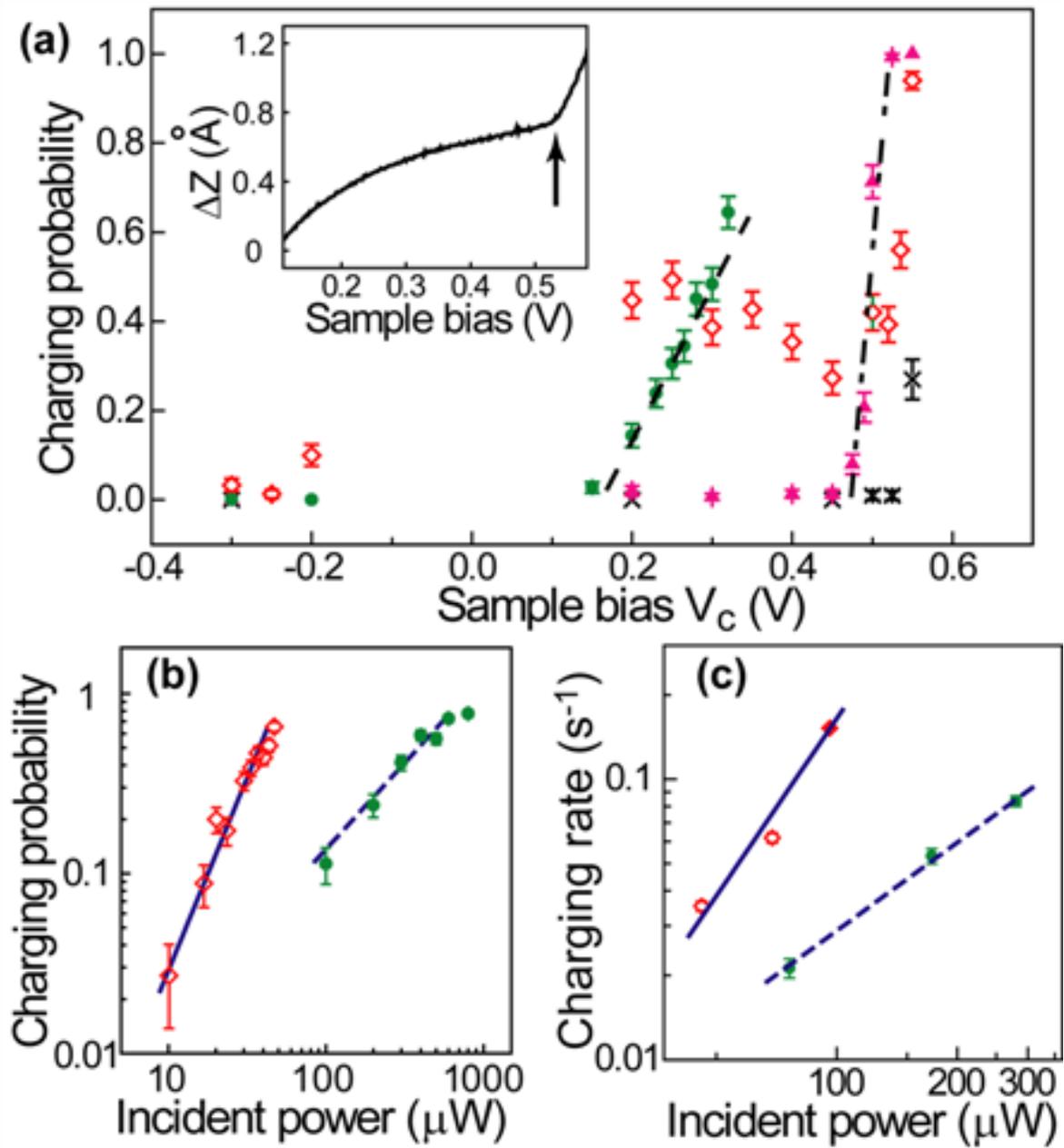

Figure 4

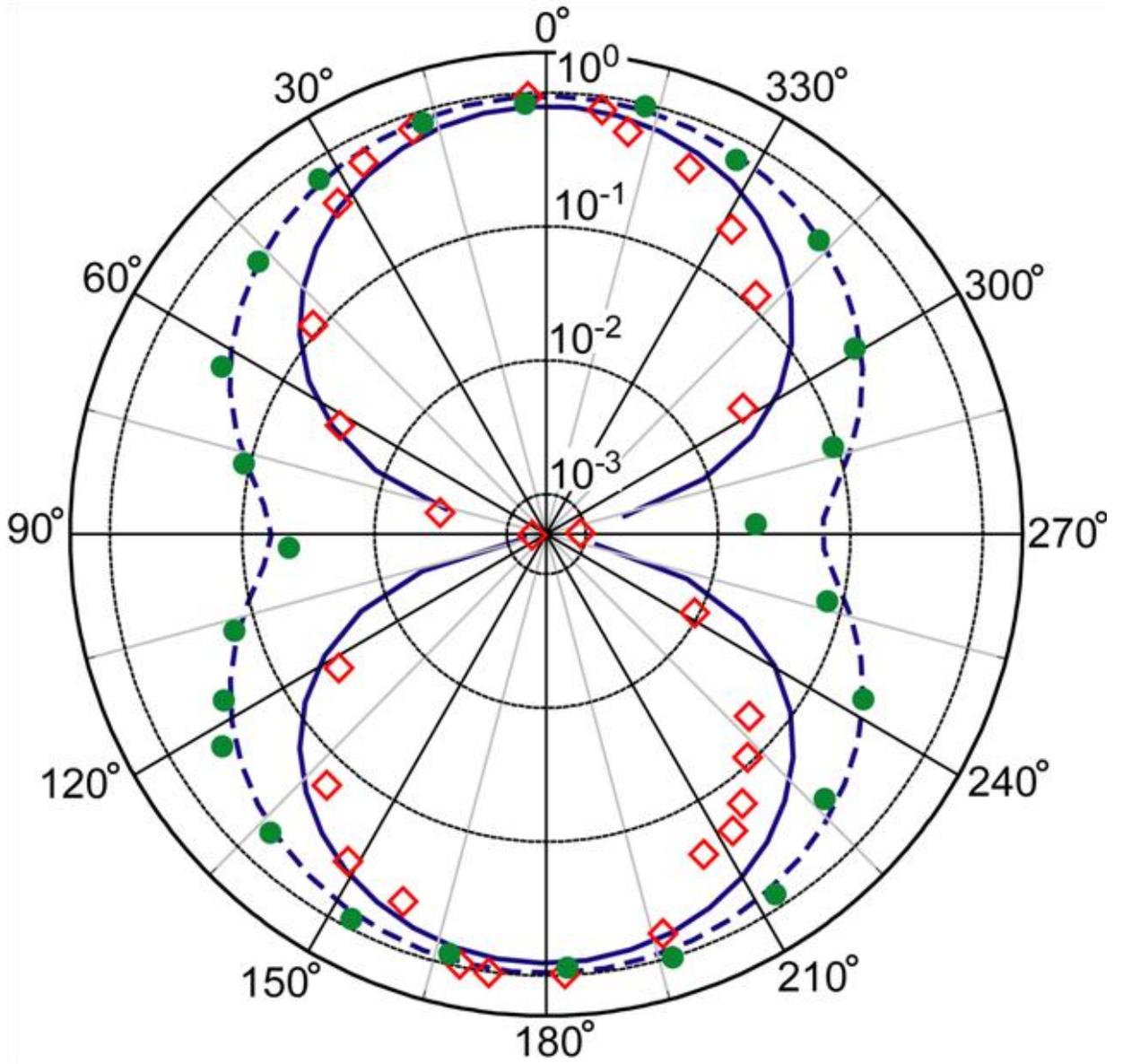

Figure 5

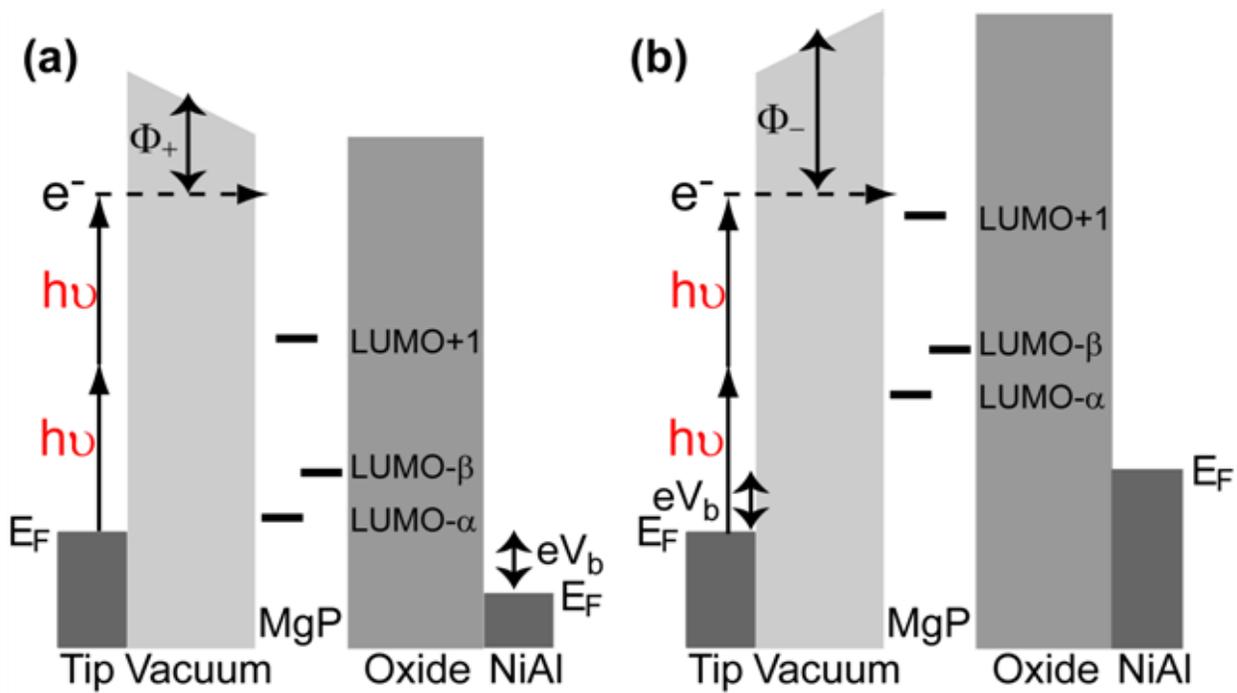

Figure 6

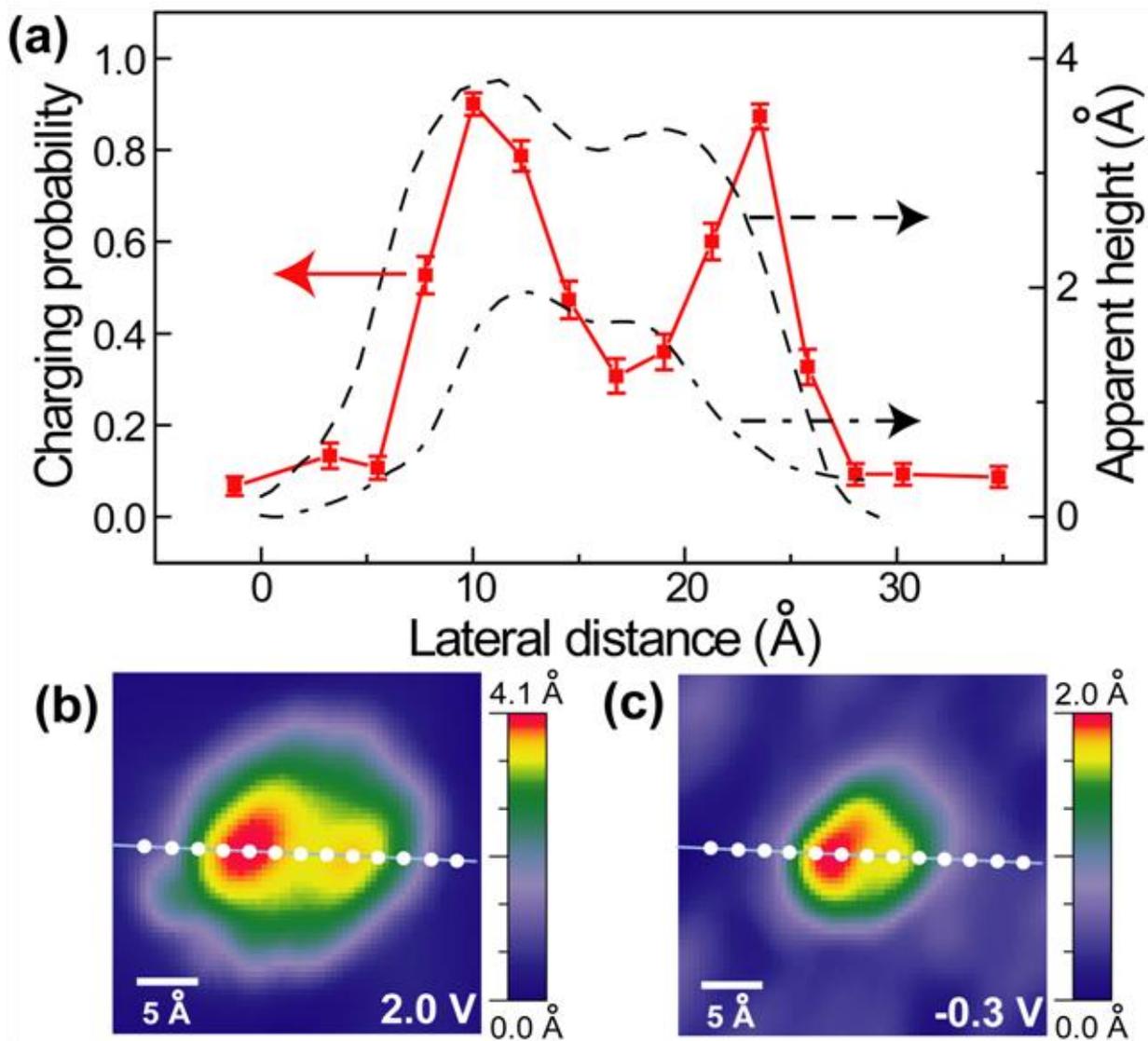

Figure 7

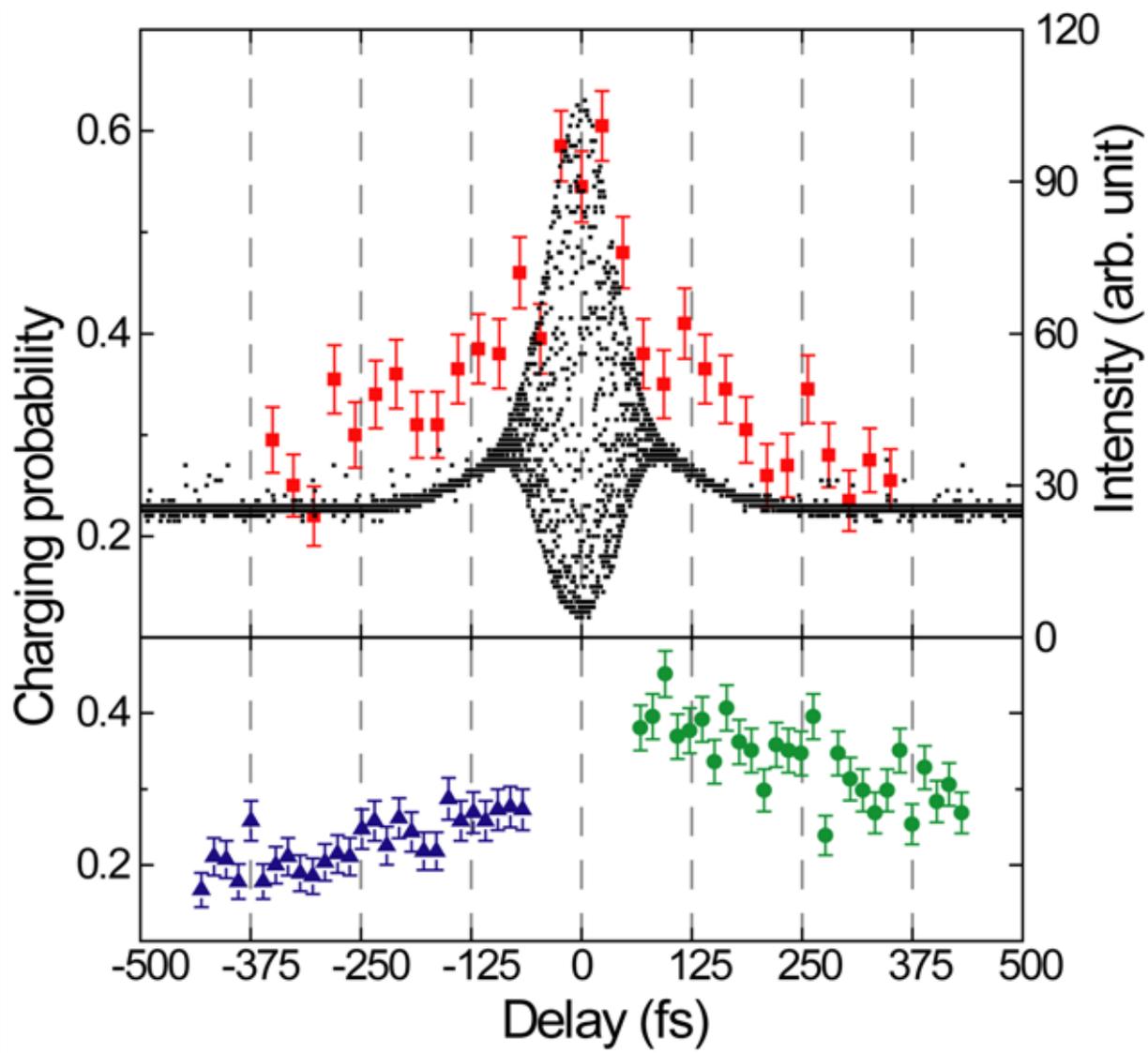

Figure 8